\documentclass{article}

\usepackage[final]{neurips_2019}

\usepackage[english]{babel}
\usepackage[utf8]{inputenc} 
\usepackage[T1]{fontenc}    
\usepackage{hyperref}       
\usepackage{url}            
\usepackage{booktabs}       
\usepackage{amsfonts}       
\usepackage{nicefrac}       
\usepackage{microtype}      
\usepackage{todonotes}
\usepackage{natbib}
\usepackage{amsmath}
\title{Give me (un)certainty - An exploration of parameters that affect segmentation uncertainty}

\author{%
  Katharina Hoebel, Ken Chang, Jay Patel, Praveer Singh, Jayashree Kalpathy-Cramer \\
   Athinoula A. Martinos Center for Biomedical Imaging\\
   Boston, MA, USA \\
   \texttt{khoebel@mit.edu} \\
}

\begin{document}

\maketitle

\begin{abstract}
  Segmentation tasks in medical imaging are inherently ambiguous: the boundary of a target structure is oftentimes unclear due to image quality and biological factors. As such, predicted segmentations from deep learning algorithms are inherently ambiguous. Additionally, "ground truth" segmentations performed by human annotators are in fact weak labels that further increase the uncertainty of outputs of supervised models developed on these manual labels. To date, most deep learning segmentation studies utilize predicted segmentations without uncertainty quantification. In contrast, we explore the use of Monte Carlo dropout U-Nets for the segmentation with additional quantification of segmentation uncertainty. We assess the utility of three measures of uncertainty (Coefficient of Variation, Mean Pairwise Dice, and Mean Voxelwise Uncertainty) for the segmentation of a less ambiguous target structure (liver) and a more ambiguous one (liver tumors). Furthermore, we assess how the utility of these measures changes with different patch sizes and cost functions. Our results suggest that models trained using larger patches and the weighted categorical cross-entropy as cost function allow the extraction of more meaningful uncertainty measures compared to smaller patches and soft dice loss. Among the three uncertainty measures Mean Pairwise Dice shows the strongest correlation with segmentation quality. Our study serves as a proof-of-concept of how uncertainty measures can be used to assess the quality of a predicted segmentation, potentially serving to flag low quality segmentations from a given model for further human review.
  
\end{abstract}

\section{Context}
\label{intro}
\selectlanguage{english}
To date, most studies on the use of deep learning to segment structures, organs or pathologies from medical imaging have treated the problem as a supervised mapping between the input image and a binary output segmentation mask without considering the inherent ambiguity of segmentation tasks, which varies depending on the imaging modality, image quality, and biological factors of the target structures. As a result, we lack absolute ground truth when we use human manual annotations, reflected by the presence of inter- and intra-rater variability \citep{Menze2015TheBRATS.}.
The idea that these ambiguous segmentation tasks require a different approach that explicitly addresses the aleatoric and epistemic uncertainties involved has been addressed previously by either providing variable outputs as segmentation hypotheses \citep{Kohl2018AImages, Baumgartner2019PHiSeg:Segmentation} or gain voxel and image level information about the uncertainty of a model's output \citep{Roy2018InherentSampling, Pan2019ProstateMeasure}. 
However, some of these approaches rely either on the availability of several segmentations per annotations for each input image or work on 2d slices instead of full 3d volumes and none has addressed how the choice of patch size, cost function, and target structure influences the segmentation uncertainty. 
Here, we explore the relationship between the aforementioned parameters and the utility of three uncertainty measures.
To quantify uncertainty information, we leverage the output of a Monte-Carlo (MC) dropout U-Net which approximates Bayesian neural networks by applying dropout after each convolutional layer at training and inference time \citep{Gal2015DropoutLearning, Kendall2015BayesianUnderstanding}.

\section{Methods}
\label{methods}

\paragraph{Data set} We use the liver and liver tumor segmentation data set provided by the medical decathlon challenge \citep{Simpson2019AAlgorithms}. For this project, we limit our examination to the 131 CT studies in the original training data set for which ground truth annotations of the liver and tumor are provided. To compare the effect and meaningfulness of uncertainty measures for less and more ambiguous segmentation tasks, we separate the two class segmentation task (segmentation of liver and tumors) into two single class segmentation tasks and train models separately for liver as well as liver tumor segmentation. The data set consisting of 131 abdominal CT scans is split randomly into training, validation, and test data sets (63/13/55).

\paragraph{Preprocessing}
For the liver segmentation task, the original scans and label maps are downsampled to 256x256x256 voxels. 
The naturally occurring differences in liver size and shape are larger than the differences in resolution caused by resampling to uniform volume size. 
The CT volumes are subsequently windowed using a soft tissue window (-120 to 240 HU) and z-score normalized based on the mean and standard deviation over the training data set. 
For the separate tumor segmentation task, we first crop the input CT to the size of a bounding box around the liver label and subsequently resample all volumes to the mean image size of the test data after cropping (284x256x133). Voxels outside of the liver mask are set to a voxel value of -50, the liver area is windowed to values between -30 and 200 to increase contrast between tumorous and non-tumorous liver tissue and normalized based on the population mean and standard deviation of the training set.

\paragraph{Segmentation models} \label{models} 
Deep learning based segmentation models are developed in DeepNeuro \citep{Beers2018}. 
Monte Carlo (MC) dropout 3d U-Nets can be seen as an approximation to Bayesian Networks by applying spatial dropout to activations during training as well as during inference \citep{Kendall2015BayesianUnderstanding, Gal2015DropoutLearning}. 
Through the application of dropout during inference we sample over segmentation outputs and can derive information about the networks uncertainty. 
The final segmentation mask is obtained by binarizing the voxelwise average over all N samples drawn for an input. 
Hereafter, we set $N = 10$. 

\paragraph{Uncertainty information}\label{unceratinty_methods}
Voxel-wise segmentation uncertainty from MC dropout models is estimated as the mean entropy over all $N=10$ samples generated by running inference on an input volume $N$ times: $U(x)= - \frac{1}{N} \sum_{i=1}^N p_i(x) \text{log}(p_i(x))$.
We extract the following uncertainty measures from MC dropout segmentation samples as described by \citet{Roy2018InherentSampling}: Coefficient of variation $ CV = \frac{\mathrm{Var}_i \left( \sum_x b(p_i(x)) \right)}{\mathrm{E}_i\left[ \sum_x b(p_i(x))\right] + 1}$, where $b(p(x))$ represents the binarized value for voxel x with the threshold $t=0.5$, mean pairwise dice over all binarized samples $D_{pw}$, and mean uncertainty over all labelled voxels $U_{labelled} = \frac{1}{\sum_{x}b(p(x))}\sum_{p(x)\geq0.5}U(x)$.  

\section{Results}
\label{results}
\subsection{Liver Segmentation}
\begin{figure} \label{fig:visual_eval}
  \centering
  \includegraphics[width=1.0\linewidth]{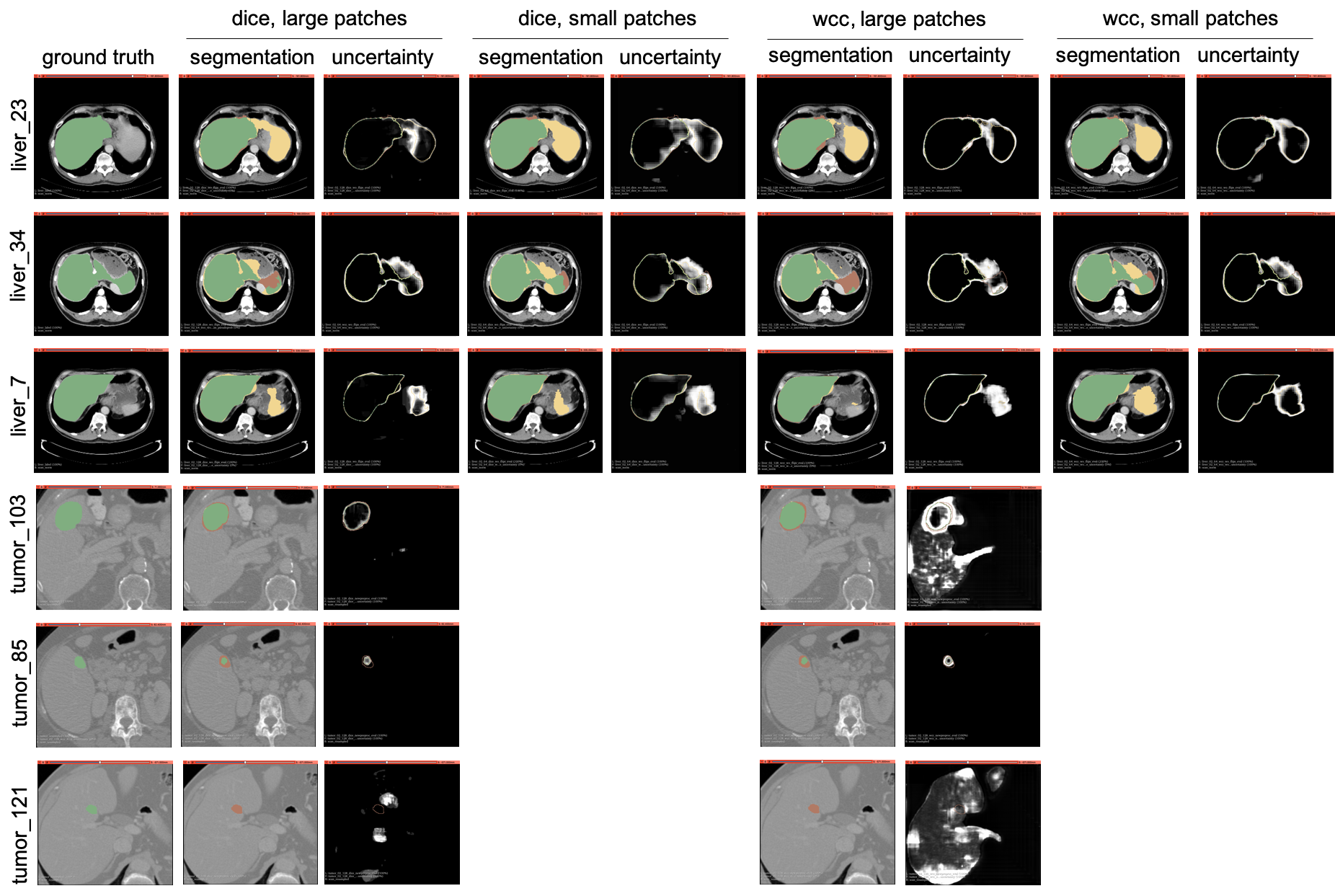}
  \caption{Segmentation predictions (green: true positive, yellow: false positive, red: false negative) and spatial uncertainty maps (brighter areas correspond to higher uncertainty) for the four developed liver and two liver tumor segmentation models.}
\end{figure}

We trained four separate Monte Carlo U-Net models for the liver segmentation task using a combination of two different cost functions (soft dice and weighted categorical crossentropy, wcc) and patch sizes (large: 128x128x16 and small: 64x64x16). 
The mean dice on the hold out test dataset ranged from 0.94 (wcc, large patches) to 0.86 (soft dice, small patches) (table  \ref{tab:liver}). \\
Examples of the spatial distribution of the voxelwise uncertainty are shown in figure \ref{fig:visual_eval}, row 1-3, column 3, 5, 7, and 9.  
On visual assessment, the voxelwise segmentation uncertainty maps of all four models show high uncertainties in the periphery of the segmentations. 
The dice U-Net uncertainty maps have a slightly steeper uncertainty gradient at the  margins as compared to the wcc U-Net  models.  
Qualitatively, cases where the binarized network output  showed  low  agreement  with  the  ground  truth revealed higher uncertainty (good example in figure \ref{fig:visual_eval}, row 2). 
Both models were prone to segment the heart and stomach. \\
Comparing the Spearman correlation coefficient between the three uncertainty measures ($CV$, $D_{pw}$, and $U_{labelled}$) (table \ref{tab:liver}) and dice score of a segmentation, the relationship is stronger for models trained with wcc vs. soft dice and larger vs. smaller patches. 
Here, mean pairwise dice is the uncertainty measure with the strongest correlation to segmentation quality whereas the coefficient of variation shows the weakest relationship.

\begin{table}[]
\caption{Liver segmentation models - Spearman correlation coefficient between uncertainty measures and segmentation quality test data}
\label{tab:liver}
    \centering
    \begin{tabular}{c|c|c|c|c|c}
         \toprule
         cost function & patch size & mean dice test (training) & CV (p-value) & $D_{pw}$ & $U_{labelled}$  \\
          \midrule
          dice & large &0.89 (0.90) & -0.50 (9.5e-05 )& 0.81 (1.2e-13) &  -0.64 (1.3e-07) \\
          dice & small & 0.86 (0.86) & -0.28 (0.038) &  0.59 (1.9e-06) & -0.50 (1.1e-04) \\
          wcc & large & \textbf{0.94}  (0.96)&  \textbf{-0.77} (5.4e-12) & \textbf{0.89} (5.6e-20) &  -0.69 (6.9e-09) \\
          wcc & small & 0.89 (0.90) & -0.53 (3.5e-05 ) & 0.81 (5.9e-14 ) & \textbf{-0.73} ( 2.1e-10 ) \\
          \bottomrule

    \end{tabular}
    \label{tab:my_label}
\end{table}

\subsection{Tumor Segmentation} 
\begin{figure} \label{fig:tumor_corr}
  \centering
  \includegraphics[width=1.0\linewidth]{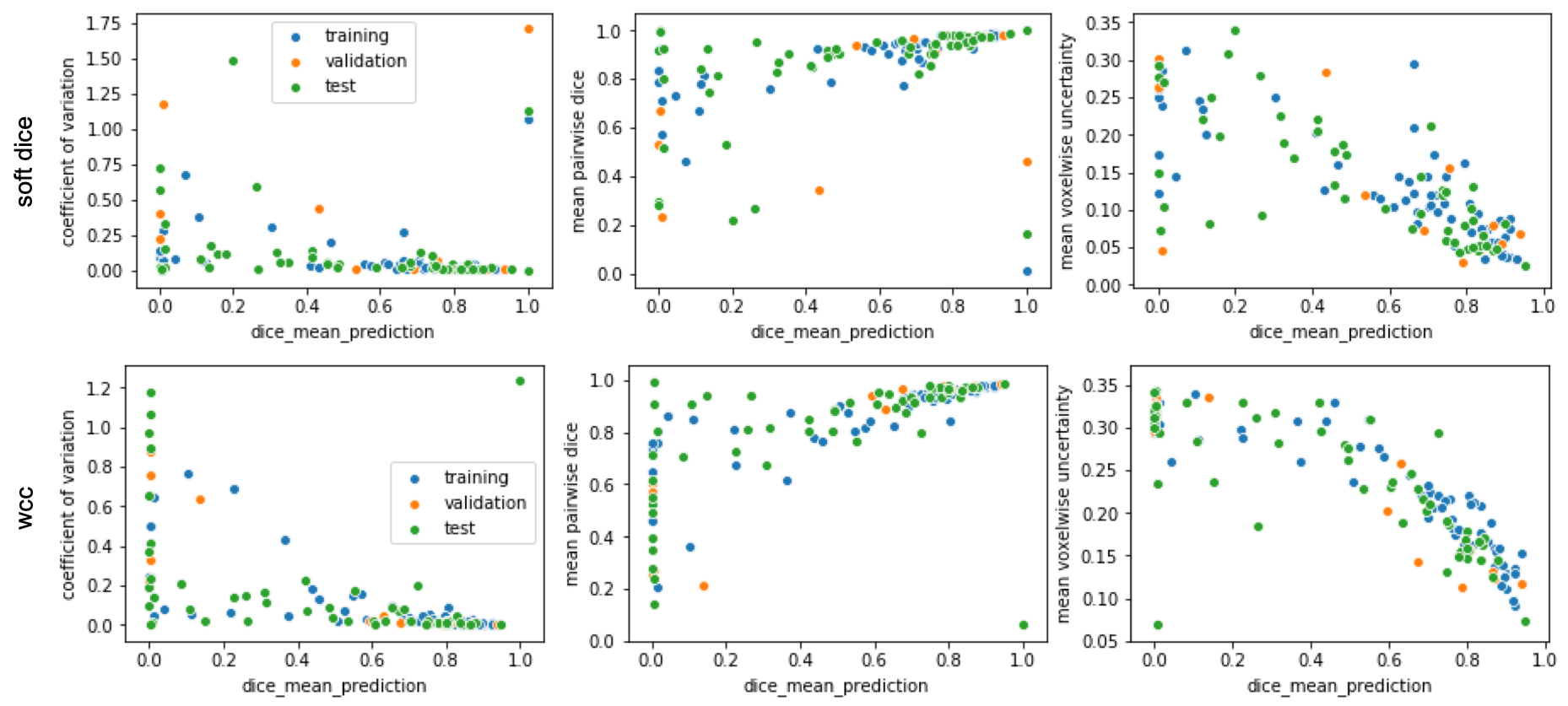}
  \caption{Correlation between segmentation quality (dice score) and uncertainty measures for the two liver tumor segmentation models for training, validation, and test data set.}
\end{figure}

For the tumor segmentation task, we trained two segmentation models (both on patches of size 128x128x16) using either soft dice or wcc as cost functions. 
The mean dice on the hold out test data set for the soft dice and wcc model is 0.52 and 0.47 respectively. 
On visual assessment the observation made on the liver segmentation uncertainty maps can be confirmed: the regions of highest uncertainty are mostly concentrated around the boundaries of the tumor in cases where it has been segmented correctly (figure \ref{fig:visual_eval} row 4 and 5).
Here, more than in the case of liver segmentation, the difference in the gradient strength of the uncertain region between dice (steep) and wcc (less steep) can be seen. 
The correlation between segmentation quality (dice score) and the three uncertainty measures are listed in table \ref{tab:tumor}. 
For all three uncertainty measures the relationship between the uncertainty measures and segmentation quality is stronger for the model trained using wcc as compared to soft dice. 
Again, even though the correlation is stronger than in the case of liver segmentation the coefficient of variation shows the weakest correlation, whereas mean labelled uncertainty shows the strongest correlation with segmentation quality with a Spearman correlation coefficient of 0.79 for the model trained using wcc. 
Figure \ref{fig:tumor_corr} shows the relationship for training, validation, and test data set.

\begin{table}[]
\caption{Tumor segmentation models - Spearman correlation coefficient between uncertainty measures and segmentation quality test data}
\label{tab:tumor}
    \centering
    \begin{tabular}{c|c|c|c|c}
         \toprule
         cost function & mean dice test (training) & CV (p-value) & $D_{pw}$ (p-value)& $U_{labelled}$ (p-value) \\
          \midrule
          dice & \textbf{0.53} (0.63) & -0.58 (7.4e-06) & 0.68 (3.4e-08) & -0.76 (1.3e-10) \\
          wcc & 0.48 (0.61) & \textbf{-0.73} (2.4e-10) & \textbf{0.79} (1.1e-12) &\textbf{ -0.79} (1.1e-12) \\
          \bottomrule

    \end{tabular}
    \label{tab:my_label}
\end{table}

\section{Conclusions}
In this exploratory work we present model uncertainty estimation derived from Monte Carlo dropout U-Nets for two different segmentation task with varying level of segmentation ambiguity: liver and liver tumor segmentation. 
Additionally, we explore the usefulness of two different cost functions, soft dice and weighted categorical cross-entropy and the influence of patch size on these uncertainty measures. 
Notably, uncertainty measures extracted from model trained using weighted categorical cross-entropy and larger patches correlate better with segmentation quality than models trained using soft dice loss and smaller patches respectively. Of the three uncertainty measures examined here, mean pairwise dice shows the strongest relationship with segmentation quality for both liver as well as liver tumor segmentation.
As the segmentation performance between the models trained for the two segmentation tasks differs greatly, we don't know how the strength of the relationship between certain uncertainty measures and segmentation quality might depend on the models performance itself and how much is caused by the ambiguity of the task. 
However, our results indicate that uncertainty measures can be used to flag failure cases for human review offering the potential to increase trust into and acceptance of automatic segmentations in clinical medicine as well as research.  
Future work should aim to develop a deeper understanding of the factors that affect segmentation uncertainty and explore the use of these measures e.g. for flagging model output for human review in large scale studies. 




\subsection*{Acknowledgements}
This publication was supported from the Martinos Scholars fund to K. Hoebel. Its contents are solely the responsibility of the authors and do not necessarily represent the official views of the Martinos Scholars fund. 

This project was supported by the National Institute of Biomedical Imaging and Bioengineering (NIBIB) of the National Institutes of Health under award number 5T32EB1680 to K. Chang and J. Patel and by the National Cancer Institute of the National Institutes of Health under Award Number F30CA239407 to K. Chang. The content is solely the responsibility of the authors and does not necessarily represent the official views of the National Institutes of Health. 

This study was supported by National Institutes of Health grants U01-CA154601, U24-CA180927, and U24-CA180918 to J. Kalpathy-Cramer. 

This research was carried out in whole or in part at the Athinoula A. Martinos Center for Biomedical Imaging at the Massachusetts General Hospital, using resources provided by the Center for Functional Neuroimaging Technologies, P41EB015896, a P41 Biotechnology Resource Grant supported by the National Institute of Biomedical Imaging and Bioengineering (NIBIB), National Institutes of Health.

\selectlanguage{english}
\bibliography{references} 

\end{document}